\documentclass[letterpaper,twocolumn,english,aps,pra,floatfix]{revtex4}
\usepackage{color}
\usepackage{verbatim}
\usepackage{amsmath}
\usepackage{amssymb}
\usepackage{graphicx}

\makeatletter

\pdfpageheight\paperheight
\pdfpagewidth\paperwidth

\@ifundefined{textcolor}{}
{%
 \definecolor{BLACK}{gray}{0}
 \definecolor{WHITE}{gray}{1}
 \definecolor{RED}{rgb}{1,0,0}
 \definecolor{GREEN}{rgb}{0,1,0}
 \definecolor{BLUE}{rgb}{0,0,1}
 \definecolor{CYAN}{cmyk}{1,0,0,0}
 \definecolor{MAGENTA}{cmyk}{0,1,0,0}
 \definecolor{YELLOW}{cmyk}{0,0,1,0}
}

\@ifundefined{definecolor}
 {\usepackage{color}}{}
\@ifundefined{definecolor}{\usepackage{color}}{}
\usepackage{times}

\renewcommand{\[}{\begin{equation}}
\renewcommand{\]}{\end{equation}}

\usepackage{babel}

\usepackage{babel}

\makeatother

\usepackage{babel}
\begin{document}
\global\long\def\avg#1{\langle#1\rangle}

\global\long\def\p{\prime}

\global\long\def\dg{\dagger}

\global\long\def\ket#1{|#1\rangle}

\global\long\def\bra#1{\langle#1|}

\global\long\def\proj#1#2{|#1\rangle\langle#2|}

\global\long\def\inner#1#2{\langle#1|#2\rangle}

\global\long\def\tr{\mathrm{tr}}

\global\long\def\pd#1#2{\frac{\partial#1}{\partial#2}}

\global\long\def\spd#1#2{\frac{\partial^{2}#1}{\partial#2^{2}}}

\global\long\def\der#1#2{\frac{d#1}{d#2}}

\global\long\def\im{\imath}

\global\long\def\As{{^{\sharp}}\hspace{-1mm}\mathcal{A}}

\global\long\def\Fs{{^{\sharp}}\hspace{-0.7mm}\mathcal{F}}

\global\long\def\Es{{^{\sharp}}\hspace{-0.5mm}\mathcal{E}}

\global\long\def\Fd{{^{\sharp}}\hspace{-0.7mm}\mathcal{F}_{\delta}}

\global\long\def\S{\mathcal{S}}

\global\long\def\A{\mathcal{A}}

\global\long\def\F{\mathcal{F}}

\global\long\def\E{\mathcal{E}}

\global\long\def\O{\mathcal{O}}

\global\long\def\SgF{\S d\F}

\global\long\def\SgEF{\S d\left(\E/\F\right)}

\global\long\def\U{\mathcal{U}}

\global\long\def\V{\mathcal{V}}

\global\long\def\H{\mathbf{H}}

\global\long\def\SO{\Pi_{\S}}

\global\long\def\PO{\hat{\Pi}_{\S}}

\global\long\def\SSH{\tilde{\Pi}_{\S}}

\global\long\def\EO{\Upsilon_{k}}

\global\long\def\ESH{\Omega_{k}}

\global\long\def\HSF{\mathbf{H}_{\S\F}}

\global\long\def\HSEF{\mathbf{H}_{\S\E/\F}}

\global\long\def\HS{\mathbf{H}_{\S}}

\global\long\def\ES{H_{\S}(t)}

\global\long\def\ESo{H_{\S}(0)}

\global\long\def\EgF{H_{\SgF} (t)}

\global\long\def\EgE{H_{\S d\E}(t)}

\global\long\def\EgEF{H_{\SgEF} (t)}

\global\long\def\EF{H_{\F}(t)}

\global\long\def\EFo{H_{\F}(0)}

\global\long\def\ESF{H_{\S\F}(t)}

\global\long\def\ESEF{H_{\S\E/\F}(t)}

\global\long\def\ESSEF{H_{\tilde{\S}\S\E/\F}(t)}

\global\long\def\EEFo{H_{\E/\F}(0)}

\global\long\def\EEF{H_{\E/\F}(t)}

\global\long\def\MI{I\left(\S:\F\right)}

\global\long\def\aMI{\left\langle \MI\right\rangle _{\Fs}}

\global\long\def\BS{\Pi_{\S} }

\global\long\def\PB{\hat{\Pi}_{\S} }

\global\long\def\QD{\mathcal{D}\left(\Pi_{\S}:\F\right)}

\global\long\def\QDp{\mathcal{D}\left(\PB:\F\right)}

\global\long\def\JI{J\left(\Pi_{\S}:\F\right)}

\global\long\def\CI{H\left(\F\left|\Pi_{\S}\right.\right)}

\global\long\def\CS{\rho_{\F\left|s\right.}}

\global\long\def\CSu{\tilde{\rho}_{\F\left|s\right.}}

\global\long\def\CSp{\rho_{\F\left|\hat{s}\right.}}

\global\long\def\ACSp{\sigma_{\F\left|\hat{s}\right.}}

\global\long\def\CEF{H_{\F\left|s\right.}}

\global\long\def\CEFp{H_{\F\left|\hat{s}\right.}}

\global\long\def\psiz{\ket{\psi_{\E\left|0\right.\hspace{-0.4mm}}}}

\global\long\def\psio{\ket{\psi_{\E\left|1\right.\hspace{-0.4mm}}}}

\global\long\def\psiinner{\inner{\psi_{\E\left|0\right.\hspace{-0.4mm}}}{\psi_{\E\left|1\right.\hspace{-0.4mm}}}}

\global\long\def\QDz{\boldsymbol{\delta}\left(\S:\F\right)_{\left\{  \sigma_{\S}^{z}\right\}  }}

\global\long\def\NQD{\bar{\boldsymbol{\delta}}\left(\S:\F\right)_{\BS}}

\global\long\def\EFS{H_{\F\left| \BS\right. }(t)}

\global\long\def\EFSM{H_{\F\left| \left\{  \ket m\right\}  \right. }(t)}

\global\long\def\Hol{\chi\left(\Pi_{\S}:\F\right)}

\global\long\def\Holp{\chi\left(\PB:\F\right)}

\global\long\def\ch{\raisebox{0.5ex}{\mbox{\ensuremath{\chi}}}_{\mathrm{Pointer}}}

\global\long\def\rhoS{\rho_{\S}(t)}

\global\long\def\rhoSo{\rho_{\S}(0)}

\global\long\def\rhoSF{\rho_{\S\F} (t)}

\global\long\def\rhoSgEF{\rho_{\SgEF} (t)}

\global\long\def\rhoSgF{\rho_{\SgF} (t)}

\global\long\def\rhoF{\rho_{\F}(t)}

\global\long\def\rhoFp{\rho_{\F}(\pi/2)}

\global\long\def\LE{\Lambda_{\E}(t)}

\global\long\def\LEc{\Lambda_{\E}^{\star}(t)}

\global\long\def\LEij{\Lambda_{\E}^{ij}(t)}

\global\long\def\LF{\Lambda_{\F}(t)}

\global\long\def\LFij{\Lambda_{\F}^{ij} (t)}

\global\long\def\LFc{\Lambda_{\F}^{\star}(t)}

\global\long\def\LEF{\Lambda_{\E/\F} (t)}

\global\long\def\LEFij{\Lambda_{\E/\F}^{ij}(t)}

\global\long\def\LEFc{\Lambda_{\E/\F}^{\star}(t)}

\global\long\def\Lkij{\Lambda_{k}^{ij}(t)}

\global\long\def\Hb{H}

\global\long\def\kE{\kappa_{\E}(t)}

\global\long\def\kEF{\kappa_{\E/\F}(t)}

\global\long\def\kF{\kappa_{\F}(t)}

\global\long\def\ts{t=\pi/2}

\global\long\def\mc#1{\mathcal{#1}}

\global\long\def\onlinecite#1{\cite{#1}}

\title{Complementarity of quantum discord and classically accessible information}

\author{Michael Zwolak}

\address{Department of Physics, Oregon State University, Corvallis, OR 97331}

\email{mpzwolak@gmail.com}

\author{Wojciech H. Zurek}

\affiliation{Theoretical Division, MS-B213, Los Alamos National Laboratory, Los
Alamos, NM 87545}

\affiliation{Santa Fe Institute, Santa Fe, NM 87501}

\maketitle
\textbf{\textcolor{black}{The sum of the Holevo quantity (that bounds
the capacity of quantum channels to transmit classical information
about an observable) and the quantum discord (a measur}}\textbf{e
of the quantumness of correlations of that observable) yields an observable-independent
total given by the quantum mutual information. This split naturally
delineates information about quantum systems accessible to observers
-- information that is redundantly transmitted by the environment
-- while showing that it is maximized for the quasi-classical pointer
observable. Other observables are accessible only via correlations
with the pointer observable. Further, we prove an anti-symmetry property
relating accessible information and discord. It shows that information
becomes objective -- accessible to many observers -- only as quantum
information is relegated to correlations with the global environment,
and, therefore, locally inaccessible. The resulting complementarity
explains why, in a quantum Universe, we perceive objective classical
reality while flagrantly quantum superpositions are out of reach.}

\textcolor{black}{There is now overwhelming evidence that the Universe
we inhabit is made out of quantum ``stuff'', and therefore quantum
to the core. This suggests that we should routinely encounter superpositions.
Yet, the world we perceive is resolutely classical. This contrast
between quantum expectations and everyday classical reality sets up
the problem that puzzled Bohr, Einstein, and many others since the
inception of quantum physics \cite{Bohr28-1,Bohr35-1,Schrodinger35-1,Einstein35-1,Bohr58-1}.}

\textcolor{black}{Decoherence \cite{Joos03-1,Zurek03-1,Schlosshauer08-1}
changed our view of the quantum-classical correspondence by explaining
the stability of pointer states that are selected in the presence
of the environment \cite{Zurek81-1,Zurek82-1}. Their nature -- in
particular, their persistence -- made them obvious candidates for
``classical states'': It was natural to expect that predictably
evolving states are good candidates for our everyday ``classical
reality''. Yet, the underlying question -- ``Why do we, observers,
perceive pointer states?'' -- remains unanswered even after recognizing
the role of decoherence in suppressing non-local superpositions. The
stability of pointer states fulfills the expectation of predictability
built on the daily experience of the classical realm, but it does
not address the obvious question: Why is it that observers choose
to measure the Universe in a way that reveals pointer states? The
key premise of this paper can be summed-up by saying that the choice
is made not by observers, but by the medium through which we perceive
the Universe.}

\begin{figure}[b]
\begin{centering}
\textcolor{black}{\includegraphics[width=8.5cm]{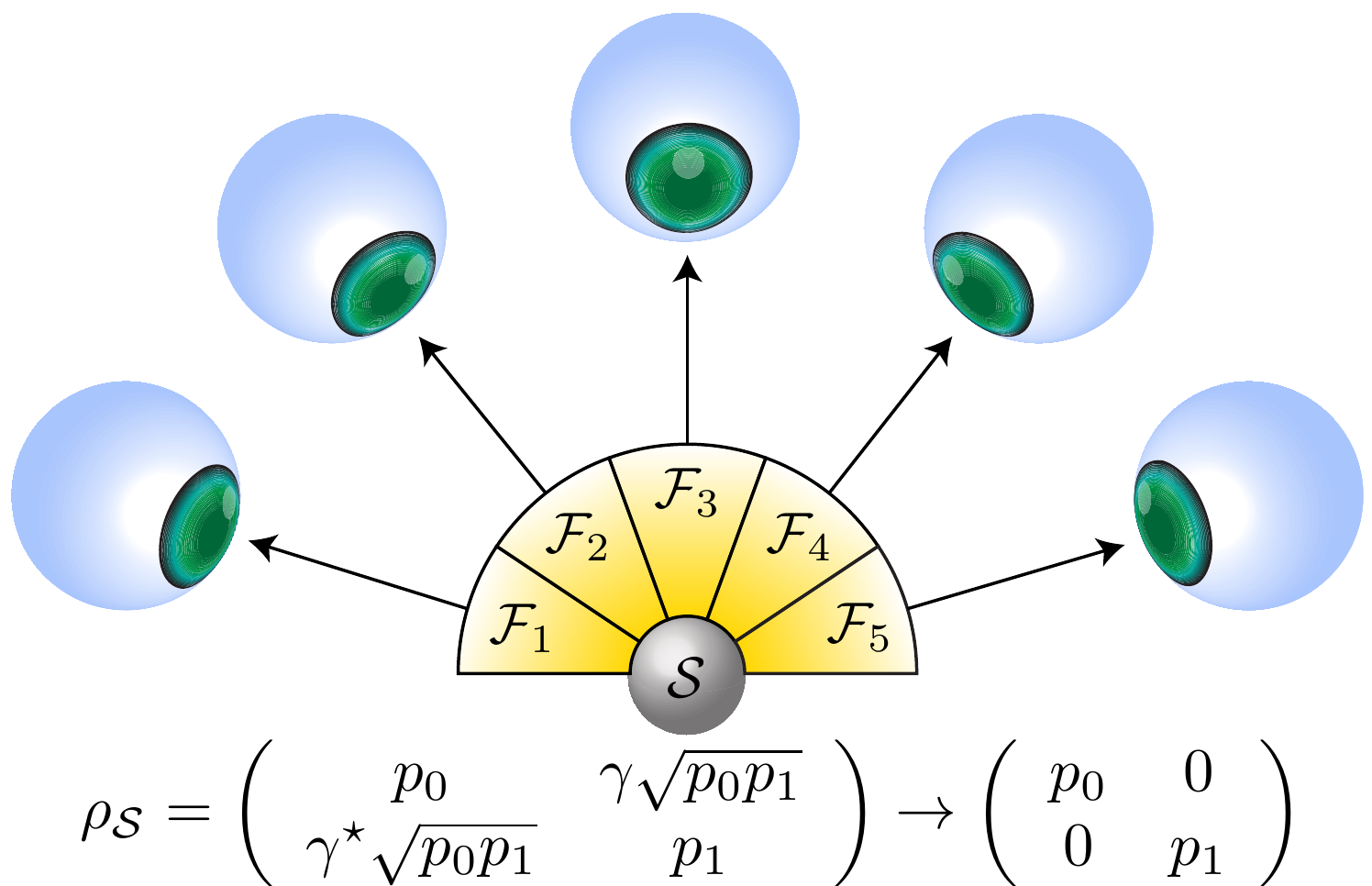} }
\par\end{centering}

\textcolor{black}{\caption{\emph{The environment as a communication channel.} A system, $\S$,
interacts with an environment, $\E$, composed of many different fragments,
$\F$. While the system decoheres (as shown by the density matrix
on the bottom), the environment fragments each acquire information
about $\S$ that can then be transmitted to observers. To learn about
the state of the system, each observer intercepts a different fragment.\label{fig:Schematic} }
}
\end{figure}

\textcolor{black}{Quantum Darwinism \cite{Ollivier04-1,Ollivier05-1,Blume05-1,Blume06-1,Blume08-1,Bennett08-1,Brunner08-1,Zwolak09-1,Zurek09-1,Paz09-1,Zwolak10-1,Riedel10-1,Burke10-1,Riedel11-1,Riedel12-1}
recognizes that the same environment that is responsible for decoherence
serves also as a channel through which information about systems reaches
observers, see Fig. \ref{fig:Schematic}. We obtain most of our data
from the photon environment. The focus of Quantum Darwinism is the
redundancy -- the presence of multiple copies -- of data about certain
observables achieved at the expense of the information about complementary
observables. Thus, the decohering environment serves not just as a
disposal for uncomfortably quantum evidence, but plays a role analogous
to a communication channel, an advertising medium in which multiple
copies of selected states of the system are present.}

\textcolor{black}{}

\textcolor{black}{Here we show that recognizing the environment as
a communication channel is far more than an allegory. Rather, it leads
to a precise split of the quantum mutual information between the system
and the environment into two components in proportions that depend
on the observable of the system: The (maximum) amount of the accessible
information about an observable is given by the Holevo quantity that
sets an upper limit on the capacity of a quantum channel to transmit
classical data \cite{Kholevo73-1}. The information that is there
in principle, but cannot be found out from the environment alone is
given by the quantum discord \cite{Zurek00-1,Ollivier02-1,Henderson01-1}
that characterizes the quantumness of correlations. The Holevo quantity
is largest for the pointer observable and decreases for other observables,
nearly vanishing for observables that are complementary. Quantum discord
makes up the difference between the mutual information (that remains
constant) and the Holevo quantity. Thus, under very general conditions
this yields a conservation law: While the classically accessible information
and quantum discord depend on the observable of the system, their
sum does not.}

\textcolor{black}{This division of the mutual information between
the Holevo quantity and the quantum discord allows one to understand
why the data about the system accessible to observers are effectively
limited to the pointer observable. We show that the Holevo quantity
for other observables decreases depending on the degree of ``misalignment''
between them and the eigenstates of the pointer observable. Furthermore,
we prove an anti-symmetry relation between discord and the Holevo
quantity. This shows that whenever objective, classical information
about a system is present, quantum information, as measured by the
discord, about this system is out of reach for observers without access
to nearly the whole environment and the system -- a situation that
can occur, at best, only in controlled laboratory experiments.}

\textcolor{black}{}
\begin{figure*}
\begin{centering}
\textcolor{black}{\includegraphics[width=16cm]{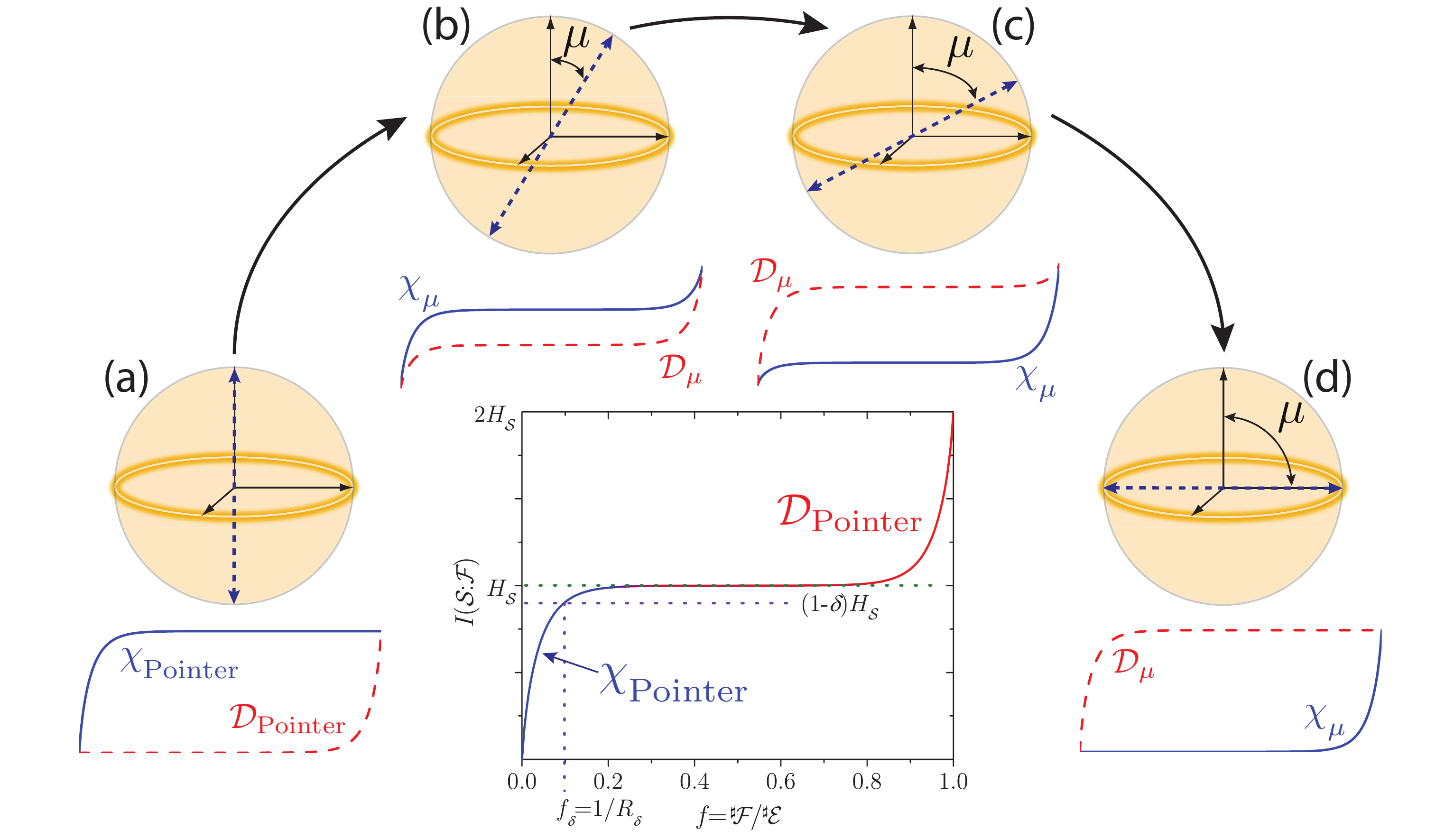} }
\par\end{centering}

\textcolor{black}{\caption{\emph{Accessible information and quantum discord.} The central plot
shows the quantum mutual information versus the size of the fragment
(given as the fraction $f=\Fs/\Es$ of the environment) for a pure
$\S\E$ state, where $\E$ has decohered $\S$. For this and the other
figures, we take the system to be two-dimensional (spin-1/2 like)
and the interaction Hamiltonian to be diagonal in the $\sigma_{z}$
basis (so the eigenstates of $\sigma_{z}$ are ``pointer''). (a)
The initial steep rise is attributable to $\chi_{\mathrm{Pointer}}$,
i.e., classical information about the pointer states of $\S$ being
communicated by the environment. Due to the anti-symmetry of the mutual
information, the purely quantum information about $\S$ is encoded
in global correlations with the environment. (b-d) Rotation by an
angle $\mu=\pi/6,\,2\pi/6,\,3\pi/6$ of the observable (i.e., the
basis given by $\ket +=\cos\left(\mu/2\right)\ket 0+\im\sin\left(\mu/2\right)\ket 1$
and $\ket -=\sin\left(\mu/2\right)\ket 0-\im\cos\left(\mu/2\right)\ket 1$)
will eventually exchange $\chi$ and $\mathcal{D}$. This gradual
change between the plot of the Holevo quantity and quantum discord
illustrate their complementary functions in the setting of the Quantum
Darwinism: Information that is locally accessible is maximized for
the pointer observable of $\S$. Accessible information about the
other observables decreases, and nearly disappears for the complementary
observable. Meanwhile, the quantum discord $\mathcal{D}$ increases,
so that the sum $\chi+\mathcal{D}$ is constant, independent of the
system observable. \label{fig:antisym}}
}
\end{figure*}

\subsection*{\textcolor{black}{Quantum Darwinism}}

\textcolor{black}{Observers typically learn about a system of interest,
$\S$, indirectly. That is, the environment $\E$ interacts with the
system and its fragments acquire information about $\S$. What the
fragment $\F$ and $\S$ know about each other is quantified by the
mutual information}

\textcolor{black}{
\[
\MI=H_{\S}+H_{\F}-H_{\S\F},
\]
which measures the total correlations present \cite{Groisman05-1}.
Here, $H$ is the von Neumann entropy of the reduced density matrices
of $\S$, $\F$, and $\S\F$. One of the main quantities of interest
in Quantum Darwinism is the typical fragment size needed by an observer
to learn about the system. That is, what typical fragment size $\Fd$
contains $1-\delta$ of the missing information $H_{\S}$ about the
system, 
\begin{equation}
\MI\ge\left(1-\delta\right)H_{\S},\label{eq:GotInfo}
\end{equation}
 decreasing the remaining entropy to $\delta\cdot H_{\S}$. The parameter
$\delta$ is the }\textit{\textcolor{black}{information deficit}}\textcolor{black}{,
which quantifies the error tolerance of the observers. All entropic
quantities that depend on the fragment are averaged with respect to
all fragments of the same size, i.e., $\MI=\aMI$. We will, on occasion,
use a definition of the fragment size $\Fd$ that replaces $\MI$
in \eqref{eq:GotInfo} with Holevo quantity $\chi$. We will show,
in the case of pointer states, that unless the fragment encompasses
almost all of $\E$, $\MI$ is essentially equivalent to $\chi$.}

\textcolor{black}{When an observer can acquire information about the
system from a small fragment of the environment, this means information
is redundant -- not only can a single observer learn about the system,
but many observers can do so independently, and hence objectively.
The redundancy is given by}

\textcolor{black}{
\[
R_{\delta}=\Es/\Fd\equiv1/f_{\delta},
\]
 where $f_{\delta}$ is the fraction size needed to satisfy Eq. \eqref{eq:GotInfo}.
The redundancy depends on how the information about $\S$ is stored
in $\E.$ Obviously, its magnitude is contingent on the information
deficit $\delta$ and the total size of the environment, $\Es$. When
many environment components independently interact with the system
-- such as photons with an object in space -- $\MI$ shown in the
central plot of Fig. \ref{fig:antisym} appears as a consequence.
The form of this curve indicates the presence of redundant information,
as indicated in the figure.}

\textcolor{black}{There is a natural connection between Quantum Darwinism
and decoherence -- in particular, the existence of pointer states.
These states survive decoherence and thus ``live on'' to proliferate
information about themselves into the environment.}

\textcolor{black}{An operational definition of pointer states (introduced
in \cite{Zurek93-1,Zurek93-2,Paz00-1} and known as the ``predictability
sieve'') is based on an intuitive idea: Pointer states can be defined
as the ones which become minimally entangled with the environment
in the course of decoherence. The predictability sieve criterion is
a way to quantify this: For every initial pure state, one measures
the entanglement generated dynamically between the system and the
environment by computing the entropy or some other measure of predictability
from the reduced density matrix of the system. The entropy is a function
of time and a functional of the initial state. Pointer states are
obtained by minimizing entropy over the initial states and demanding
that the answer be robust when varying the decoherence time.}

\textcolor{black}{Pointer states are important in determining what
information is deposited in the environment. In addition to this connection,
the main themes of this work will be how $\MI$ naturally separates
into classical and quantum components, and the implications for the
emergence of the classical world.}

\section{\textcolor{black}{Results}}

\subsection*{\textcolor{black}{The Holevo Quantity and Discord}}

\textcolor{black}{We start with a straightforward rewrite of the definition
of quantum discord in the setting suitable for Quantum Darwinism:
We consider a system $\S$ that is decohered by the environment $\E$.
We focus on a fragment $\F$ of $\E$. Quantum discord (from $\S$
to $\F$) is then defined as 
\begin{equation}
\QD=\MI-\JI\label{eq:Discord}
\end{equation}
 for the POVM $\BS$. The asymmetric mutual information $\JI$ is
given by 
\begin{equation}
\JI=H_{\F}-\CI.\label{eq:AsymMI}
\end{equation}
 The conditional entropy $\CI$ depends on the density matrices $\CS$
of the fragments $\F$ conditioned on the outcomes $s$ for the POVM.
The asymmetric mutual information is equal to $\MI$ when Bayes' rule
relating joint and conditional probabilities holds, as it does for
classical systems \cite{Cover06-1}.}

\textcolor{black}{As a result of decoherence the system correlates
with the environment. For a two dimensional system,}

\textcolor{black}{
\begin{equation}
\left[\alpha\ket 0_{\S}+\beta\ket 1_{\S}\right]\ket{\psi_{\E}}\mapsto\alpha\ket 0_{\S}\psiz+\beta\ket 1_{\S}\psio,\label{eq:GenericBranch}
\end{equation}
 where $\psiz$ and $\psio$ are the conditional states of the environment
generated by interaction with the system. The extent of the coherence
between $0$ and $1$ states of the system will depend on the overlap
between the corresponding states of $\E$, $\psiinner$. In the course
of decoherence, the environment also acquires a record of the system's
state. Orthogonal conditional states on the environment perfectly
record the state of the system. Thus, there is a correspondence between
decoherence and the acquisition of a record. As we will see below,
in common decoherence scenarios such records are redundant and nearly
complete when $0,1$ is the pointer basis. Moreover, the environment
then acts as a communication channel, broadcasting the record into
the larger world.}

\textcolor{black}{This is not just an analogy. The asymmetric mutual
information in Eq. \eqref{eq:AsymMI} is the well-known Holevo quantity
\[
\Hol=H\left(\sum_{s}p_{s}\CS\right)-\sum_{s}p_{s}\CEF,
\]
 where $p_{s}$ is the probability of outcome $s$ occurring (see
Methods). The Holevo quantity bounds the amount of classical information
transmittable over a quantum channel, i.e., the classically accessible
information. In the example above, this information is about the states
$0$ or $1$ (more generally, the POVM $\BS$). With the recognition
that $J$ and $\chi$ are identical, we can now write 
\begin{equation}
\MI=\Hol+\QD.\label{eq:Conservation}
\end{equation}
 This is the conservation law that we now employ in the discussion
of Quantum Darwinism. It is illustrated in Fig. \eqref{fig:antisym}.
Its salient feature is the fact that its left hand side does not depend
on what is of interest to an observer via the set $\BS$, while the
ingredients -- the classical and quantum components -- on the right
hand side do.}

\textcolor{black}{In the above discussion, we have implicitly assumed
that arbitrary POVMs are allowed. In most prior discussions of quantum
discord, $\BS$ is usually (although not always) taken as a set of
orthogonal states. However, Datta \cite{Datta08-2} has shown that
quantum discord can always be minimized by using rank one projectors.
We accept this generalization to arbitrary POVMs.}

\textcolor{black}{In the study of discord there is also a natural
temptation to extremize (usually, minimize) $\QD$ with respect to
the set $\BS$. We are interested in what happens to $\chi$ and $\mathcal{D}$
as the measurement characterized by $\BS$ is varied. This requires
a departure from the usual temptations and a consistent -- but more
general -- interpretation of discord.}

\textcolor{black}{The sharp division of the whole information present
in the $\S\E$ correlations into classical, locally accessible $\chi$
and quantum $\mathcal{D}$ (that can be accessed only globally by
measurements involving both $\S$ and $\E$) puts immediately to rest
the concern that was raised in the early discussions of decoherence
\cite{Greenberger89-1,Carmichael94-1}, where the common criticism
was that ``one has to ignore the environment to justify the emergence
of the pointer observable''. Our discussion demonstrates that --
by virtue of Holevo's theorem -- no measurement of the environment
alone can reveal more than $H_{\S}$ -- the missing part of the information
about the quasi-classical pointer states $\S$. Only global measurement
of $\S$ and all of $\E$ can detect phase coherence that is a quantum
``leftover'' from the phase coherence in the initial state of $\S$.}

\textcolor{black}{Thus, classicality that emerges from decoherence
does not rest on the assumption of ignoring the environment, but on
the realization that the measurements available to the observer are
local -- that they do not involve global observables with eigenstates
that are entangled states of $\S$ and the whole of $\E$. Indeed,
even this (already outlandish) requirement does not suffice: To reveal
global coherence one would have to measure the state of $\S\E$ in
such a way that it would not collapse -- i.e., one would have to choose
a priori a global observable that has that (unknown) entangled state
as an eigenstate. A number of well-known facts (including, e.g., the
no cloning theorem \cite{Wootters82-1,Dieks82-1}) make this impossible.}

\subsection*{\textcolor{black}{Anti-symmetry and the emergence of classicality}}

\textcolor{black}{Let us now analyze the consequences of this split
into quantum and classical information. Given an arbitrary rank one
POVM and a pure, but otherwise arbitrary, state $\rho_{\S\E}$, the
quantum discord and the Holevo quantity on complementary fragments
of the environment (i.e., $\F$ and the rest of $\E$, $\E/\F$) are
related by 
\begin{eqnarray}
\mathcal{D}\left(\Pi_{\S}:\E/\F\right) & = & I\left(\S:\E/\F\right)-\chi\left(\Pi_{\S}:\E/\F\right)\nonumber \\
 & = & H_{\S}-H_{\S\E/\F}+\sum_{s}p_{s}H_{\E/\F\left|s\right.}\nonumber \\
 & = & H_{\S}-H_{\F}+\sum_{s}p_{s}\CEF\nonumber \\
 & = & H_{\S}-\Hol\label{eq:Equal}
\end{eqnarray}
 where we started with Eq. \eqref{eq:Conservation}. The second to
last line used that, for a globally pure state, the conditional state
of $\E$ with respect to a rank one POVM element is also pure. The
latter implies $H_{\E/\F\left|s\right.}=\CEF$ since $\E/\F$ and
$\F$ is a bipartite split of the conditional state of $\E$, just
as $H_{\S\E/\F}=H_{\F}$. }

\textcolor{black}{The relation, Eq. \eqref{eq:Equal}, implies that
when classical information about $\BS$ is available from small fragments
of the environment, $\Hol\ge H_{\S}\left(1-\delta\right)$, then quantum
information is banished to global correlations with the environment,
$\mathcal{D}\left(\Pi_{\S}:\E/\F\right)\le\delta\cdot H_{\S}$. Therefore,
local observers -- even when they can intercept the rest of the environment,
$\E/\F$ -- will not have access to it and will never detect superpositions.
The reverse is also true for pure $\S\E$ states -- when quantum information
is contained in global correlations, then classical information will
be present in small fragments of the environment.}

\textcolor{black}{}
\begin{figure}
\begin{centering}
\textcolor{black}{\includegraphics[width=8.5cm]{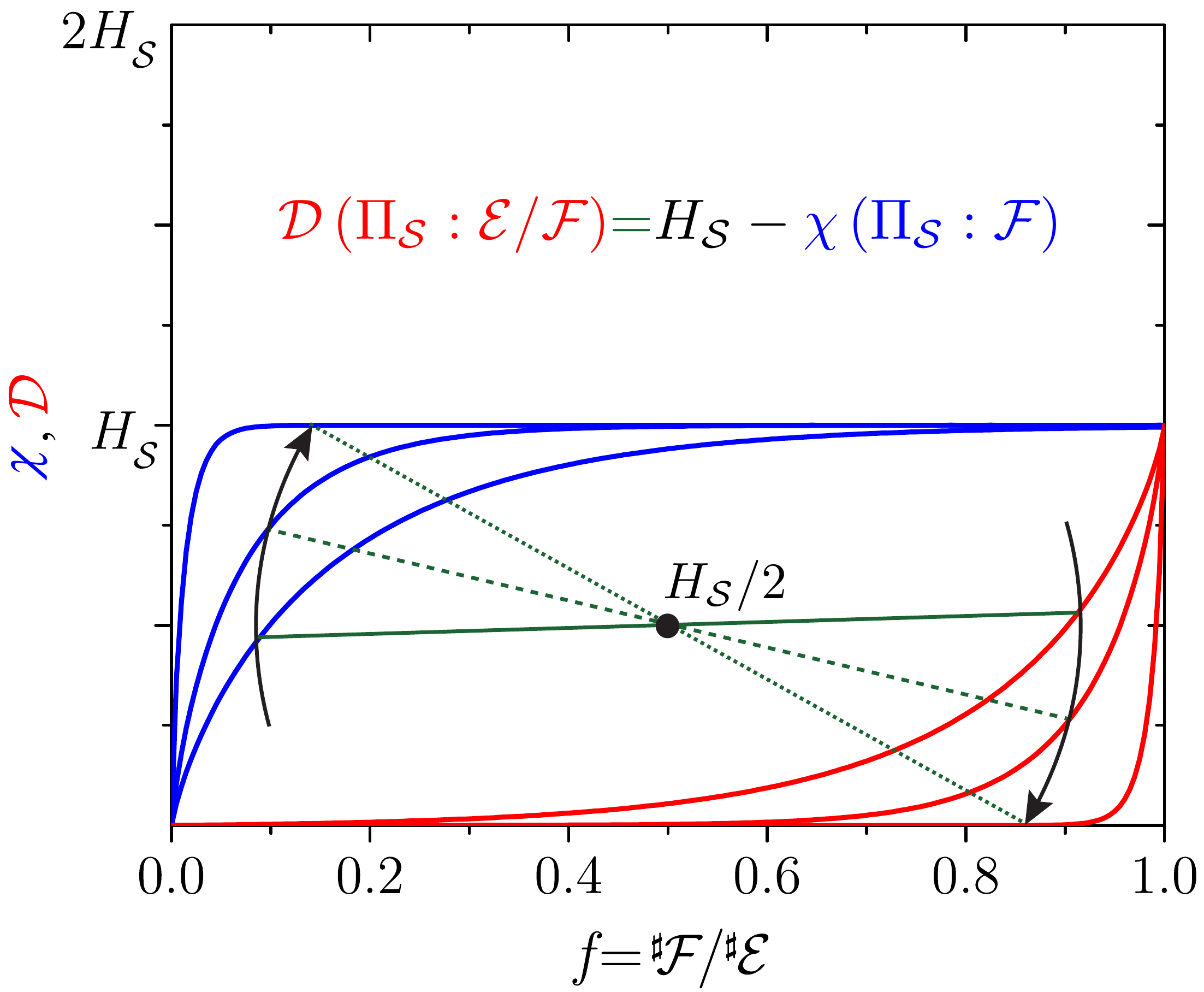} }
\par\end{centering}

\textcolor{black}{\caption{\emph{Anti-symmetry of discord and the Holevo quantity for a pure
$\S\E$ state and rank one POVM $\Pi_{\S}$}. This symmetry relates
discord between $\Pi_{\S}$ and the partial environment $\E/\F$ and
the Holevo quantity between $\Pi_{\S}$ and fragment $\F$. As the
Holevo quantity increases -- i.e., as the classical information transmitted
by the environment increases -- the discord on the opposite side of
the axis decreases. Since the discord is monotonically increasing
with $f$, this means that the quantum information about $\S$ is
pushed into correlations with the global environment as redundant
(classical) information is increased. Thus, a state which has redundant
information will have quantum information encoded in the environment,
implying that Quantum Darwinism is in action. \label{fig:Anti-symmetry}}
}
\end{figure}

\textcolor{black}{It is worth reflecting on this statement further.
Figure \ref{fig:Anti-symmetry} shows the results of Eq. \eqref{eq:Equal}
rewritten to pivot about $H_{\S}/2$: 
\[
\Hol-H_{\S}/2=H_{\S}/2-\mathcal{D}\left(\Pi_{\S}:\E/\F\right).
\]
 This shows that increasing the redundant (classical) information
stored in a small fragment $\F$ decreases the quantum information
in the much larger fragment $\E/\F$. In other words, a world where
objective information is present is also a world with quantum information
inaccessible to all but the most encompassing observer. The loss or
inaccessibility of any tiny component of the environment will preclude
the retrieval of this quantum information.}

\textcolor{black}{For arbitrary $\S\E$ states, including mixed states,
Eq. \eqref{eq:Equal} is replaced by the inequality (see Methods)
\[
\mathcal{D}\left(\Pi_{\S}:\E/\F\right)\leq H_{\S}-\Hol.
\]
Thus, it can be stated unequivocally that whenever redundant information
is present, quantum information is relegated to global correlations
with the environment. We can, as well, examine what happens in particular
mixed states. If only $\S$ is mixed, for instance, the discord will
just be further suppressed -- not only is it contained solely in global
correlations, but it can be totally absent (see, e.g., Ref. \cite{Zwolak10-1}).
In the cases studied in Refs. \cite{Zwolak09-1,Zwolak10-1}, the initial
mixedness of $\E$ reduced the redundancy but quantum information
stayed banished to global correlations.}

\subsection*{\textcolor{black}{Branching States and Surplus Decoherence}}

\textcolor{black}{To ``set the baseline'' for physical application
of these results, we simplify the rest of our study to the two scenarios
that accurately approximate commonly encountered situations -- the
emergence of branching states via decoherence \cite{Blume05-1,Blume06-1,Everett57-1}
and what we will call ``surplus decoherence''. Branching happens
when the evolution of the decohering system does not create transitions
between its pointer states, and the imprints of these states on the
environment components are unaffected by the evolution of the remaining
$\E$. This is a good approximation of what happens when the photon
environment scatters from a heavy object: Subsystems of the environment
(photons) individually interact with the system, each pushing $\S$
closer to a localization, but they do not interact with each other
\cite{Riedel10-1,Riedel11-1}. Fragments $\F$ consist of collection
of such subsystems. On a more formal level, the situation will give
rise to Eq. \eqref{eq:GenericBranch} where $\ket{\psi_{\E\left|s\right.}}=\bigotimes_{k}\ket{\psi_{k\left|s\right.}}$.
Branching states can also occur approximately, at times shorter than
the dissipation timescale \cite{Blume08-1} or the mixing timescale
of environment components \cite{Riedel12-1}.}

\textcolor{black}{Surplus decoherence occurs when a part of $\E$
suffices to decohere $\S$ -- i.e., when both the environment $\E$
and the environment without some fragment $\E/\F$ completely decohere
the system. In other words, the environment is so large that the state
of the system and a fragment will have the form 
\begin{equation}
\rho_{\S\F}=\sum_{\hat{s}}p_{\hat{s}}\proj{\hat{s}}{\hat{s}}\otimes\CSp,\label{eq:GoodDec}
\end{equation}
 where $\CSp$ is the conditional state of the fragment given the
state $\hat{s}$. That is, there is some orthogonal basis $\PB$ of
the system -- the pointer basis -- such that $\E/\F$ will decohere
even the joint state of $\S\F$ giving a discord-free form. In many
situations (but not always) branching will give rise to surplus decoherence.
Together, these two cases accurately approximate commonly encountered
system-fragment states generated by decoherence. Thus, it is enlightening
to study what are the consequences of these conditions for the division
of the mutual information into classical and and quantum components.}

\textcolor{black}{For the final part of the paper, we will mostly
make use of pure, branched $\S\E$ states, which have the form 
\begin{equation}
\ket{\psi_{\S}}\left[\bigotimes_{k\in\E}\ket{\psi_{k}}\right]\mapsto\sum_{\hat{s}}c_{\hat{s}}\ket{\hat{s}}\left[\bigotimes_{k\in\E}\ket{\psi_{k\left|\hat{s}\right.}}\right],\label{eq:Branched}
\end{equation}
 where $\hat{s}$ indicates a pointer state. For these states, $\MI$
has the shape characteristic of Quantum Darwinism, see Fig. \ref{fig:antisym}:
It raises steeply from $\MI=0$ at $\Fs=0$, the size of $\F$, to
a plateau of $H_{\S}$ -- to the level of the entropy of the system.
This plateau is in many ways the dominant feature of $\MI$ that arises
in Quantum Darwinism.}

\subsubsection*{\textcolor{black}{Pointer states minimize discord}}

\textcolor{black}{When $\MI$ is decomposed into $\chi$ and $\mathcal{D}$,
the initial rise is attributable to the channel capacity $\chi$ that
reveals the state of the pointer basis. The mutual information continues
then at the level $H_{\S}$ until $\Fs\simeq\Es$ where the fragment
becomes so large that it encompasses essentially all of the environment.
For a state of the form in Eq. \eqref{eq:Branched}, the discord is
\begin{eqnarray}
\QD & = & H_{\S}-H_{\S\F}+\sum_{s}p_{s}H_{\F\left|s\right.}\nonumber \\
 & = & \QDp+\sum_{s}p_{s}H_{\F\left|s\right.},\label{eq:DisBranchState}
\end{eqnarray}
 where the second line follows from the fact that the states of $\F$
conditional on a pointer state on $\S$ are pure (just a pure $\E$
state will suffice to ensure the conditional state on $\F$, $\CSp$,
is pure). The second term in Eq. \eqref{eq:DisBranchState} is positive
but zero for the pointer basis and thus the pointer basis minimizes
the discord. Using the conservation law, when the discord is minimized,
the Holevo quantity is maximized. Therefore, the maximum accessible
quantum information is largest for the pointer basis.}

\textcolor{black}{Moreover, for branched states, including ones evolved
from initially mixed states of $\S\E$, the discord is given by $\QDp=H_{\S d\E}-H_{\S d\E/\F}$,
i.e., a difference of the system entropy decohered by the full environment
$\E$ and by the partial environment $\E/\F$ (see Eq. (17) of Ref.
\cite{Zwolak10-1}). For initially mixed states, the pointer basis
will not necessarily minimize the discord. However, the discord with
respect to the pointer basis is exponentially small in the size of
the environment, and we have the state approximately given by Eq.
\eqref{eq:GoodDec}. Thus, for all practical purposes, one has 
\[
\MI=\Holp.
\]
 This equality is equivalent to the assumption of surplus decoherence.
While the surplus decoherence condition can be framed independently
of the dynamics/Hamiltonian, we know that the physically relevant
scenario of many environment components interacting independently
with the system (e.g., the photon environment) leads naturally to
surplus decoherence. If one removed a small fragment of the photon
environment, systems would still be rapidly decohered. In fact, in
this scenario, Eq. \eqref{eq:GoodDec} is true up to corrections $\mathcal{O}\left(\prod_{k\in\E/\F}\Lambda_{k}\right)$,
where $\left|\Lambda_{k}\right|<1$ is the decoherence factor from
environment component $k$. Yet, the state resulting from surplus
decoherence has a deep implication: The mutual information between
$\S$ and $\F$ contains only classical information about the pointer
basis, $\MI=\chi\left(\PB:\F\right)$. }

\textcolor{black}{For a pure state of $\S$ and $\E$, the resulting
$\MI$ is antisymmetric with respect to $H_{\S}$ and $\Fs=\Es/2$.
This means that the initial rise to the plateau must be matched by
the steep rise from the plateau to $I\left(\S:\F\to\E\right)=2H_{\S}$
as the size of $\F$ approaches the size of the whole of $\E$. This
final rise is caused by the rapid increase of $\QDp$ from $0$ to
$H_{\S}$.}

\textcolor{black}{The physical implications of these runs of $\chi$
and $\mathcal{D}$ are straightforward and appealing: The information
that can be obtained from the environment about the pointer observable
of $\S$ by measurement of a fragment $\F$ quickly saturates to the
value set by its entropy $H_{\S}$. Moreover, only $\chi$ of information
can be obtained from $\E$ no matter how large is the fragment $\F$.
This follows form Holevo's theorem \cite{Kholevo73-1}: $\chi$ is
an upper bound on the information that can be extracted from the quantum
channel. The rapid rise of $\MI$ at the very end is then completely
due to discord -- due to the quantum information that can be accessed
only via global measurements that involve both $\S$ and $\E$.}

\textcolor{black}{This consideration implies that when many environment
components interact independently with the system, only classical
information will be transferred into the environment: There is a basis
chosen by the environment's interaction with the system that is proliferated
into the environment. In this sense, the generation of branching states
will always proliferate information about the pointer basis. Incidentally,
this is the world in which we live, where photons interact independently
with systems, proliferating redundant -- and therefore objective --
information and conveniently hiding quantum information. We can take
this latter step forward, and show that one can only -- in a way that
will be clear in a moment -- find out about the pointer states.}

\subsubsection*{\textcolor{black}{Only pointer states can be redundant}}

\textcolor{black}{We now consider an attempt to extract, from fragments
of $\E$, information about $\sigma_{y}$, which is complementary
to the pointer observable $\sigma_{z}$ (see Fig. \ref{fig:antisym}).
Success would imply detection of evidence of quantumness -- catching
Schr\"odinger's cat in a superposition of dead and alive. The plot of
quantum mutual information $\MI$ is, of course, independent of the
observable of $\S$. The two contributions, though, ``change places''.
Now it is the discord that raises rapidly (a feature that we address
below), its graph matching the plot of $\chi$ in the previous pointer
observable case. By contrast, $\chi$ remains close to zero until
the very end, where measurement of all of $\E$ could in principle
reveal the eigenstate of $\sigma_{y}$ with which $\E$ is entangled.
Still, there is no information about $\sigma_{y}$ that can be gleaned
from any fraction of $\E$ -- the whole of $\E$ is needed to get
$\sigma_{y}$ (if it is available at all).}

\textcolor{black}{}
\begin{figure*}
\begin{centering}
\textcolor{black}{\includegraphics[width=16cm]{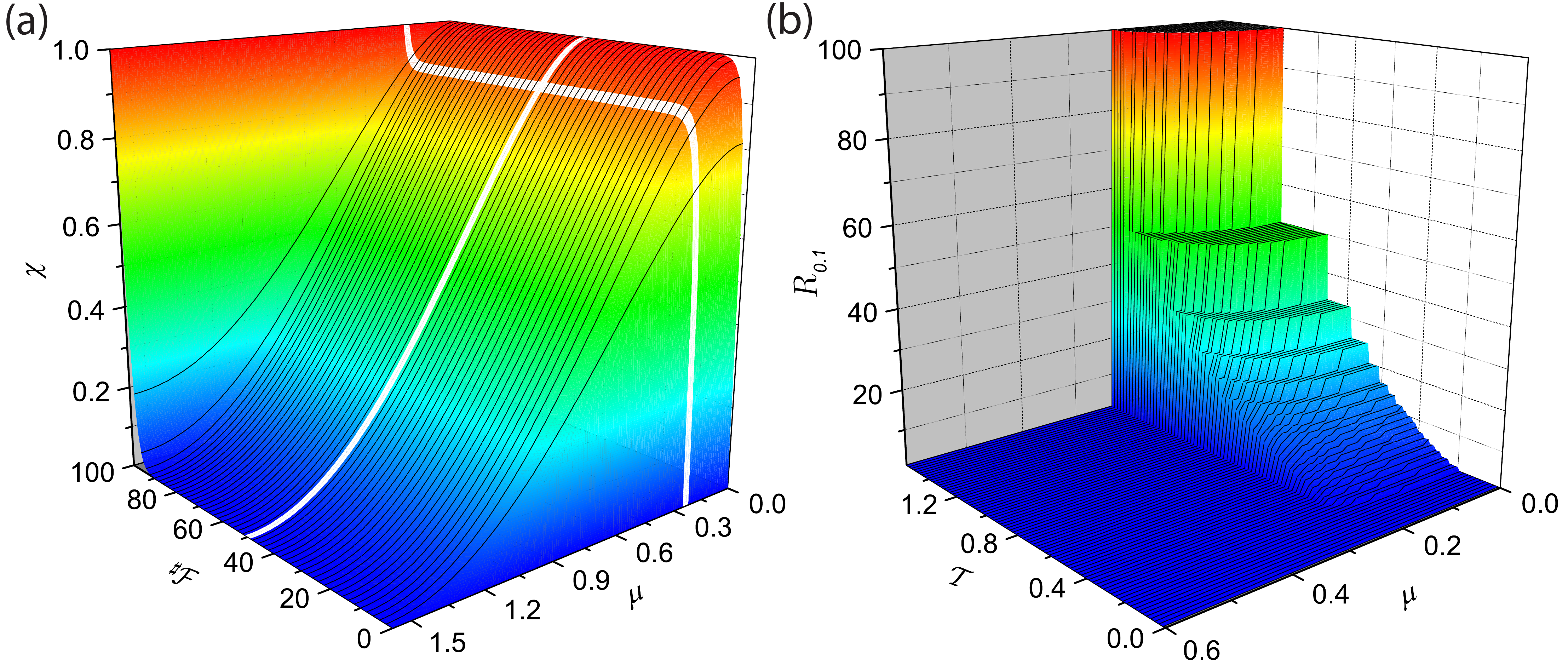} }
\par\end{centering}

\textcolor{black}{\caption{\emph{Observable dependence of the classical information} \emph{and
its redundancy}. (a) Plot of the Holevo quantity versus basis (defined
by the angle $\mu$) and fragment size for the representative spin
model. Here, the off-diagonal elements of the reduced state of the
system, $\rho_{\S}$, are suppressed by the decoherence factor $\left(\cos\mathcal{T}\right)^{\Es}$,
where $\mathcal{T}$ is the action (in units of $\hbar$) that results
from the coupling of a single environment spin to the system. The
plot shows the Holevo quantity for $\mathcal{T}=\pi/2$, $\Es=100$,
and a symmetric and pure environment state. The white line parallel
to the fragment size axis demarcates where redundancy is no longer
possible for $\delta=0.1$. The other white line plots Eq. \eqref{eq:ChiPlat}
for the plateau value of $\chi$. Note that the discord versus basis
would be a similar figure, but larger on the side $\mu=\pi/2$, as
it is given by $\QD=\MI-\Hol$. (b) Redundancy versus $\mu$ and $\mathcal{T}$
computed by finding $\Fd$ such that $\Hol\ge H\left(\BS\right)\left(1-\delta\right)$.
Compared to Eq. \eqref{eq:GotInfo}, this calculation of $\Fd$ drops
terms that are exponentially small. For the example case here, the
missing information is $H\left(\BS\right)=H\left(p_{\mu}\right)=1$,
with $H\left(p_{\mu}\right)$ the binary entropy of the new probability
distribution. The probabilities of detecting the spin in the ``+/-''
directions along the axis defined by $\mu$ are $p_{+}=p_{0}\cos^{2}\left[\mu/2\right]+p_{1}\sin^{2}\left[\mu/2\right]$
and $p_{-}=1-p_{+}$, where $p_{0,1}=1/2$ are the probabilities of
the pointer states occurring\textcolor{black}{. The conditional entropy
is given by $H\left(\BS\left|\PB\right.\right)=H\left(\cos^{2}\left[\mu/2\right]\right)$.
}\label{fig:ChiAndRed} }
}
\end{figure*}

\textcolor{black}{An intermediate case -- spin at an angle $\mu$
from $\sigma_{z}$ -- is an obvious next case to consider. Now the
plateau of the corresponding $\chi_{\mu}=\Hol$ is lower than the
missing information about the ``intermediate'' observable, $H\left(\BS\right)$.
Under surplus decoherence and when $\F$ holds a perfect record of
the pointer states, $\chi$ for another observable is given by 
\begin{equation}
\Hol=H\left(\BS\right)-H\left(\BS\left|\PB\right.\right).\label{eq:ChiPlat}
\end{equation}
 Figure \ref{fig:ChiAndRed} shows the Holevo quantity versus the
fragment size and the angle for the example system under consideration.
When $H\left(\SO\left|\PB\right.\right)>\delta\cdot H\left(\BS\right)$,
redundant records of the observable $\BS$ can not exist -- at most
``there can be only one'' copy. When $\delta$ is small, but non-zero,
only states that are very nearly the pointer states can be determined.
An equivalent result to Eq. \eqref{eq:ChiPlat} was obtained in Ref.
\cite{Ollivier05-1}. We show in the Methods how to extend it to imperfect
records.}

\subsection*{\textcolor{black}{The Holevo quantity for pointer $\PB$ is the discord
for a complementary $\BS$}}

\textcolor{black}{Continuing from the above results, as $\chi_{\mu}$
is reduced, $\mathcal{D}_{\mu}$ must make up the difference due to
the conservation law. In case of surplus decoherence, this means 
\[
\Holp=\Hol+\QD.
\]
 In the extreme case -- when the basis $\BS$ is complementary to
the pointer basis $\PB$ -- the plateau in the Holevo quantity is
zero, and the accessible information about the pointer basis turns
into the complementary discord. We define a complementary POVM $\BS$
as one where all elements satisfy 
\[
\bra{\hat{s}}\pi_{s}\ket{\hat{s}}=q_{s},
\]
 where $q_{s}$ is independent of the pointer state $\hat{s}$. That
is, each $\pi_{s}$ is unbiased with respect to any pointer state
$\ket{\hat{s}}$. An example would be the $\sigma_{y}$ basis when
$\sigma_{z}$ is pointer.}

\textcolor{black}{Assuming a state of the form in Eq. \eqref{eq:GoodDec},
the conditional state with respect to the outcome $s$ is 
\begin{align}
p_{s}\CS & =\tr_{\S}\sqrt{\pi_{s}}\rho_{\S\F}\sqrt{\pi_{s}}\nonumber \\
 & =q_{s}\rho_{\F},\label{eq:Condl}
\end{align}
 yielding $p_{s}=q_{s}$. In other words, the conditional states are
just the reduced states of $\F$. Then, however, 
\begin{eqnarray*}
\QD & = & H_{\S}-H_{\S\F}+\sum_{s}p_{s}\CEF\\
 & = & H_{\S}-\left(H_{\S}+\sum_{\hat{s}}p_{\hat{s}}\CEFp\right)+\sum_{s}p_{s}H_{\F}\\
 & = & H_{\F}-\sum_{\hat{s}}p_{\hat{s}}\CEFp\\
 & = & \Holp,
\end{eqnarray*}
 where the second line follows from the surplus decoherence condition
and also Eq. \eqref{eq:Condl}. Thus, the plot of $\QD$ will follow
that of $\Holp$. When one does not exactly have surplus decoherence,
the corrections will depend on the magnitude of the off-diagonal elements
in $\rho_{\S\F}$. For many environment components independently decohering
the system these elements become exponentially small in the environment
size (exponentially small in $\Es-\Fs$).}

\section{\textcolor{black}{Discussion}}

\textcolor{black}{We have demonstrated that the quantum mutual information
naturally separates into classical and quantum components. We proved
that a world containing redundant information about a system necessarily
implies that quantum information is inaccessible. In other words,
if information about some basis has been proliferated into the environment
(making it ``objective''), then the quantum information -- information
about superpositions of that basis (e.g., of the initial state of
the system), is suppressed. This information cannot be obtained by
observers intercepting only a fragment of the environment. Rather,
they need the whole environment and the system to retrieve it. Furthermore,
the information proliferated into the environment is not destroyed
by further change/monitoring of the system. Once the system has deposited
multiple copies of its information, they are ``here to stay''.}

\textcolor{black}{Although discord is often studied these days, questions
remain about its fundamental significance. This work suggests that
discord has a robust role to play in defining what quantum information
means. We have explored it in a specific context of the quantum-to-classical
transition in the setting of Quantum Darwinism, but some of our conclusions
are clearly relevant more generally. Our results also help draw a
distinction between the more flagrant aspects of the ``quantumness
of correlations'' captured by entanglement and the quantumness of
correlations that are separable -- devoid of entanglement -- and yet
not completely classical.}

\section{\textcolor{black}{Methods\label{sec:Methods}}}

\subsection*{\textcolor{black}{The Holevo Quantity and Discord}}

\textcolor{black}{The quantum discord (from $\S$ to $\F$) is 
\begin{equation}
\QD=\MI-\JI\label{eq:DiscordSupp}
\end{equation}
 given the POVM $\BS$, $\sum_{s}\pi_{s}=\mathrm{I}$, where $\pi_{s}$
are the elements of $\BS$. The quantum mutual information is $\MI=H_{\S}+H_{\F}-H_{\S\F}$,
where $H$ is the von Neumann entropy of the of the density matrices
$\rho_{S}$, $\rho_{\F}$, and $\rho_{\S\F}$ that one obtains by
tracing out ``the rest'', i.e., $\rho_{\S}=\tr_{\E}\rho_{\S\E}$,
$\rho_{\F}=\tr_{\S\E/\F}\rho_{\S\E}$, $\ldots$. All entropic quantities
are averaged with respect to all fragments of the same size, e.g.,
$\MI=\aMI$. Most results hold without averaging, and it will be stated
when averaging is necessary.}

\textcolor{black}{The state of $\F$ conditional on outcome $s$ on
$\S$ is 
\[
p_{s}\CS=\tr_{\S}\sqrt{\pi_{s}}\rho_{\S\F}\sqrt{\pi_{s}},
\]
 where $p_{s}$ is the probability of obtaining outcome $s$. Denoting
the entropy of $\F$ given outcome $s$ as $\CEF=-\tr\CS\log\CS$,
the asymmetric mutual information is 
\begin{eqnarray}
\JI & = & H_{\F}-\CI\nonumber \\
 & = & H_{\F}-\sum_{s}p_{s}\CEF.\nonumber \\
 & = & H\left(\sum_{s}p_{s}\CS\right)-\sum_{s}p_{s}\CEF.\label{eq:Asym}
\end{eqnarray}
 The last line used 
\begin{eqnarray}
\sum_{s}p_{s}\CS & = & \sum_{s}\tr_{\S}\sqrt{\pi_{s}}\rho_{\S\F}\sqrt{\pi_{s}}\nonumber \\
 & = & \sum_{s}\tr_{\S}\pi_{s}\rho_{\S\F}=\rho_{\F},\label{eq:rhoFPOVM}
\end{eqnarray}
 which follows from the cyclic property of the partial trace (with
the identity acting on $\F$). Even though the POVM does not uniquely
set the post-measurement state of $\S$, it will for $\F$. Equation
\eqref{eq:Asym} is of course the equation for the Holevo quantity
$\chi$, which we denote as $\Hol$ since it depends on the POVM $\BS$.
The similarity with the Holevo quantity was also noticed in Ref. \cite{Henderson01-1}.}

\subsection*{\textcolor{black}{Anti-symmetry and the emergence of classicality}}

\textcolor{black}{Assuming a pure $\S\E$ state, the mutual information,
$\aMI$ (designated by $\MI$), is antisymmetric about the point $\Fs=\Es/2$
\cite{Blume06-1}. That is, 
\begin{equation}
\MI=2H_{\S}-I\left(\S:\E/\F\right)\label{eq:Antisym}
\end{equation}
 and $I\left(\S:\E/2\right)=H_{\S}$. This is shown readily by writing
out the mutual information 
\begin{eqnarray*}
\MI+I\left(\S:\E/\F\right) & = & H_{\S}+H_{\F}-H_{\S\F}\\
 &  & +H_{\S}+H_{\E/\F}-H_{\S\E/\F}\\
 & = & H_{\S}+H_{\F}-H_{\E/\F}\\
 &  & +H_{\S}+H_{\E/\F}-H_{\F}\\
 & = & H_{\S}+H_{\S}\\
 & = & H_{\S}+H_{\E}-0=I\left(\S:\E\right)
\end{eqnarray*}
 where we used that $\S\E$ is in a pure state to relate entropies
across a bipartite split (e.g., $H_{\S\F}=H_{\E/\F}$). Equation \eqref{eq:Antisym}
holds regardless of whether averaging is done. However, only when
the mutual information refers to the average of all fragments of a
given size does it imply that the mutual information versus $\Fs$
is antisymmetric about its midpoint (at $\Fs=\Es/2$, $\MI=H_{\S}$
from Eq. \eqref{eq:Antisym} when averaged). This is easy to see by
examining, e.g., the mutual information in the state 
\[
\left(\ket{00}_{\S\F_{1}}+\ket{11}_{\S\F_{1}}\right)\ket{0\cdots0}_{\E/\F_{1}},
\]
 where only a single environment spin is correlated with the system.
If ordering is maintained in the environment spins, the mutual information
is clearly not anti-symmetric, even though Eq. \eqref{eq:Antisym}
holds. It becomes so only when averaging is performed and the mutual
information is a curve versus fragment size only.}

\textcolor{black}{In the main text, we derived a stronger result 
\begin{equation}
\mathcal{D}\left(\Pi_{\S}:\E/\F\right)=H_{\S}-\Hol\label{eq:EqualSupp}
\end{equation}
 for an arbitrary rank one POVM $\BS$ and a pure state of $\S\E$.
Averaging is not required for Eq. \eqref{eq:EqualSupp}. More generally,
\begin{equation}
\mathcal{D}\left(\Pi_{\S}:\E/\F\right)\leq H_{\S}-\Hol\label{eq:SuppofDiscord}
\end{equation}
 for an arbitrary rank one POVM $\BS$ and arbitrary, potentially
mixed, state of $\S\E$. To show this, consider any state $\rho_{\S\E}$
and purify the state to $\proj{\psi_{\S\E\E^{\p}}}{\psi_{\S\E\E^{\p}}}$.
Starting with the conservation law, 
\begin{eqnarray*}
\mathcal{D}\left(\Pi_{\S}:\E/\F\right) & = & H_{\S}-H_{\S\E/\F}+\sum_{s}p_{s}H_{\E/\F\left|s\right.}\\
 & = & H_{\S}-H_{\F\E^{\p}}+\sum_{s}p_{s}H_{\F\E^{\p}\left|s\right.},\\
 & = & H_{S}-\chi\left(\Pi_{\S}:\F\E^{\p}\right)
\end{eqnarray*}
 where we used that for a rank one POVM the conditional state of $\E\E^{\p}$
is pure, and therefore $H_{\E/\F\left|s\right.}=H_{\F\E^{\p}\left|s\right.}$.
From the data processing inequality \cite{Nielsen00-1}, if we trace
out the purifying environment $\E^{\p}$, then this reduces the Holevo
quantity 
\begin{equation}
\Hol\leq\chi\left(\Pi_{\S}:\F\E^{\p}\right).\label{eq:ChiRed}
\end{equation}
 and gives Eq. \eqref{eq:SuppofDiscord}. We note that, in addition
to pure states, there are cases of interest here where equality holds
in Eq. \eqref{eq:ChiRed} (i.e., when $\F$ is mixed but only within
disjoint subspaces that are correlated with $\S$). }%

\subsection*{\textcolor{black}{Branching States and Surplus Decoherence}}

\subsubsection*{\textcolor{black}{Only pointer states can be redundant}}

\textcolor{black}{We have seen that the pointer states minimize discord
for pure, branching states (and for surplus decoherence), and that
branching states lead to surplus decoherence up to exponentially small
corrections. Let's see what happens to transmitted information for
surplus decoherence. Consider the case where a perfect record of the
pointer basis exist in a fragment of the environment, 
\begin{equation}
\Holp=H_{\S},\label{eq:HolPoint}
\end{equation}
 i.e., that the conditional states $\CSp$ are orthogonal for the
PVM $\PB$. Now consider a POVM $\BS$, that we can define by the
quantities 
\[
p_{s\hat{s}}=\bra{\hat{s}}\pi_{s}\ket{\hat{s}},
\]
 which give the conditional probabilities for $s$ to occur given
$\hat{s}$. For the state $\rho_{\S\F}$ in Eq. \eqref{eq:GoodDec},
the probability of outcome $s$ is 
\[
p_{s}=\tr_{\S\F}\pi_{s}\rho_{\S\F}=\sum_{\hat{s}}p_{s\hat{s}}p_{\hat{s}}.
\]
 Further, since the $p_{s\hat{s}}$ are conditional probabilities,
they obey 
\[
\sum_{s}p_{s\hat{s}}=1,
\]
 which is readily obtained by using that $\BS$ is a POVM. This also
gives 
\[
\sum_{s}p_{s}=1
\]
 for the probabilities for $s$ to occur.}

\textcolor{black}{The Holevo quantity for the communication of (classical)
information about $s$ is given by 
\[
\Hol=H\left(\rho_{\F}\right)-\sum_{s}p_{s}\CEF,
\]
 where we used Eq. \eqref{eq:rhoFPOVM}. Taking Eq. \eqref{eq:GoodDec}
and the assumption that for the pointer basis the classical information
is at the plateau value, i.e., that there is a perfect record, one
obtains 
\begin{eqnarray}
\Hol & = & H\left(\sum_{\hat{s}}p_{\hat{s}}\CSp\right)\nonumber \\
 &  & -\sum_{s}p_{s}H\left(\sum_{\hat{s}}p_{s\hat{s}}p_{\hat{s}}\CSp/p_{s}\right)\nonumber \\
 & = & H\left(p_{\hat{s}}\right)+\sum_{\hat{s}}p_{\hat{s}}\CEFp\nonumber \\
 &  & -\sum_{s}p_{s}\left[H\left(p_{s\hat{s}}p_{\hat{s}}/p_{s}\right)+\sum_{\hat{s}}p_{s\hat{s}}p_{\hat{s}}/p_{s}\CEFp\right]\nonumber \\
 & = & H\left(p_{\hat{s}}\right)+\sum_{\hat{s}}p_{\hat{s}}\CEFp\nonumber \\
 &  & -\sum_{s}p_{s}H\left(p_{s\hat{s}}p_{\hat{s}}/p_{s}\right)-\sum_{\hat{s}}p_{\hat{s}}\CEFp\nonumber \\
 & = & H\left(p_{\hat{s}}\right)+\sum_{s,\hat{s}}p_{s\hat{s}}p_{\hat{s}}\log\left(p_{s\hat{s}}p_{\hat{s}}/p_{s}\right)\nonumber \\
 & = & H\left(\BS\right)-H\left(\BS\left|\PB\right.\right)\label{eq:HolForGoodDec}
\end{eqnarray}
 for the plateau value of the classical information about $s$ deposited
in the environment: Its value is suppressed by the conditional entropy
of $\BS$ with respect to the pointer basis $\PB$, i.e., the misalignment
of $\BS$ from $\PB$.}

\textcolor{black}{The presence of redundant information requires that
\[
\Hol\ge H\left(\BS\right)(1-\delta),
\]
 where $H\left(\BS\right)$ measures the amount of missing information.
This can not be satisfied when 
\[
H\left(\BS\left|\PB\right.\right)>\delta\cdot H\left(\BS\right),
\]
 which makes it clear that when the POVM is rotated away from the
pointer basis, eventually redundant information cannot be present
in the environment. Furthermore, even for an environment with an infinite
number of components, whether or not there is redundant information
about the POVM depends on the accuracy required: All POVMs that are
not the pointer basis will not be redundantly encoded if one requires
a small information deficit. In this sense, we say that only pointer
states can be found out.}

\textcolor{black}{Of course, one can also start from the expression
Eq. \eqref{eq:DisBranchState} for branched states and use the conservation
law to get 
\begin{equation}
\Hol=\Holp-\sum_{s}p_{s}\CEF.\label{eq:ChiForPureDec}
\end{equation}
 Thus, the Holevo quantity will be reduced for all POVMs that are
not the pointer basis. As the generation of the branched state gives
rise to surplus decoherence and the fragment starts to acquire a perfect
record, Eq. \eqref{eq:ChiForPureDec} will approach Eq. \eqref{eq:HolForGoodDec}. }

We can quantify this approach for imperfect records. We assume surplus
decoherence and that the pointer states are correlated with nearly
distinguishable states $\CSp$, i.e., that one is very near the classical
plateau. The latter implies that there exists a POVM $\Lambda$ on
$\F$, with elements $\Lambda_{\hat{s}}$, such that $\tr\Lambda_{\hat{s}}\CSp\geq1-\epsilon$
and $\tr\Lambda_{\hat{s}^{\p}}\CSp\leq\epsilon$ for $\hat{s}^{\p}\neq\hat{s}$,
where $\epsilon$ gives the error probability for distinguishing the
states. Let the measurement be carried out by apparatus $\A$. We
can apply this to studying the Holevo quantity with respect to some
arbitrary POVM on $\S$:
\[
\Hol=\chi\left(\BS:\F\A\right)_{\rho}=\chi\left(\BS:\F\A\right)_{\tilde{\rho}},
\]
where first $\rho_{\S\F}\to\rho_{\S\F\A}=\rho_{\S\F}\otimes\ket 0_{\mathcal{A}}\bra 0$
and then a unitary acts on $\F\A$, with its relevant action defined
by $\mathcal{U}_{\F\A}\ket{\psi}_{\F}\ket 0_{\A}=\sum_{\hat{s}}\sqrt{\Lambda_{\hat{s}}}\ket{\psi}_{\F}\ket{\hat{s}}_{\A}$,
to get $\tilde{\rho}_{\S\F\A}$, neither of which change the Holevo
quantity. Then, however, the positive operator $\Omega=\sum_{\hat{s}}\proj{\hat{s}}{\hat{s}}\otimes\mathrm{I}_{\F}\otimes\ket{\hat{s}}_{\mathcal{A}}\bra{\hat{s}}\le\mathrm{I}$
has the property
\[
\tr\,\Omega\tilde{\rho}_{\S\F\A}\ge1-\epsilon.
\]
Thus, the state 
\[
\sigma_{\S\F\A}=\frac{\sqrt{\Omega}\tilde{\rho}_{\S\F\A}\sqrt{\Omega}}{\tr\,\Omega\tilde{\rho}_{\S\F\A}}
\]
has 
\[
\frac{1}{2}\tr\left|\sigma_{\S\F\A}-\tilde{\rho}_{\S\F\A}\right|\leq\sqrt{\epsilon}
\]
by the gentle measurement lemma \cite{Winter99-1,Ogawa07-1}. The
analysis leading to Eq. \eqref{eq:HolForGoodDec} holds as $\S$ and
$\A$ are perfectly correlated in the state $\sigma_{\S\F\A}$. We
can then apply the Alicki-Fannes' inequality \cite{Alicki04-1} 
\begin{align*}
\left|\Hol-\left(H\left(\BS\right)-H\left(\BS\left|\PB\right.\right)\right)\right|\\
=\left|\chi\left(\BS:\F\A\right)_{\tilde{\rho}}-\chi\left(\BS:\F\A\right)_{\sigma}\right|\\
=\left|H\left(\left.\BS\right|\F\A\right)_{\tilde{\rho}}-H\left(\left.\BS\right|\F\A\right)_{\sigma}\right|\\
\leq8\sqrt{\epsilon}\log D_{\Pi}+2H\left(2\sqrt{\epsilon}\right),
\end{align*}
where $D_{\Pi}$ is the number of outcomes for the POVM $\BS$. Thus,
the reduction in the classical information about non-pointer $\BS$
is reduced by their correlation with the pointer states up to small
corrections. For the case of branching states, these corrections are
exponentially small in the size of the fragment.

The surplus decoherence assumption can be similarly relaxed, i.e.,
by assuming $\rho_{\S\F}$ is near to the discord-free state. For
branching states, $\rho_{\S\F}$ is exponentially close in the size
of environment to a discord free state. Thus, the $\epsilon$ above,
while small, dominates how close the exact $\Hol$ is to the perfect
record case. The case of imperfect records but mixed states extends
the results of Ref. \cite{Ollivier05-1}.
\begin{acknowledgments}
This research is supported in part by the U.S. Department of Energy
through the LANL/LDRD Program and, in part, by the John Templeton
Foundation. 
\end{acknowledgments}
\bibliographystyle{naturemag}
\bibliography{QCcorr}

\appendix

\section*{Supplemental Information\label{sec:Supplemental}}

\textcolor{black}{The discord is normally interpreted as information
that is accessible only globally (i.e., in the scenario here, this
means measuring $\S$ and the whole of $\E$). This requires, however,
that one minimizes discord with respect to $\BS$. The discord without
this minimization has a more general meaning: Discord quantifies the
loss of correlations when measurement of a basis occurs. That is,
ignoring the apparatus that measures $\BS$, the correlations between
$\S$ and $\F$ are only classical and given by $\Hol=\Holp-\QD$.
This meaning is fully consistent with the previous interpretation,
but without the requirement that discord is minimized with respect
to the chosen basis. We emphasize, however, that despite the destruction
of correlations, the information on the prior state of $\S$ -- before
interaction with measurement apparatus -- is still present in $\E$
even though the correlations have been severed. }

\textcolor{black}{Before discussing this further and for the case
of rank one POVMs, we also note that in some cases one can access
the quantum information without resorting to simultaneous measurement
of both $\S$ and all of $\E$. For instance, if one wants to discriminate
the GHZ states $\left(\ket{0\cdots0}\pm\ket{1\cdots1}\right)/\sqrt{2}$,
this can be done by one at time (e.g., bit by bit) measurements on
all the qubits. What is important, however, is that one has access
to all the qubits and measures all of them (i.e., a global measurement,
but not a simultaneous measurement). If any one of them is lost or
otherwise inaccessible, then the phase information is lost. Furthermore,
you have to know ahead of time the phase information you are trying
to distinguish between. This is in contrast to trying to figure out
the redundant information, i.e., the 0's and 1's, which is a robust
process with respect to the observer's prior knowledge.}

Let us now discuss discord with respect to $\BS$ in more detail.
A natural way to interpret the conservation law, Eq. \eqref{eq:Conservation},
is to imagine decohering the $\S\F$ state in some basis $\BS$: The
state becomes 
\begin{equation}
\rho_{\S\F}=\sum_{s=1}^{D_{\S}}p_{s}\proj ss\otimes\CS,\label{eq:rhoSFmu}
\end{equation}
 where the $p_{s}$ are the probabilities to find the state $\ket s$
and $\CS$ are the conditional states on the fragment. This state
now contains only $\Hol$. The part of the state $\rho_{\S\F}$ that
was ``ignored'' -- i.e., the contributions that are off diagonal
in the basis $\BS$ -- is the information represented by the discord.
Further, the mutual information does not depend on the basis. Thus,
the information is there somewhere, but can not be accessed in $\F$
after a measurement of the $\BS$ basis has been done. This is a reflection
of the quantum nature of $\S$: A basis can be defined that is the
superposition of the states that are actually correlated with $\F$,
and this new basis shuffles classical information into ``coherences''
in the new basis.

We can discuss this more concretely by assuming surplus decoherence
in the pointer basis $\PB$, represented by the state in Eq. \eqref{eq:GoodDec}.
With a state of this form, only classical information will be present
in the fragment: 
\[
\MI=\Holp.
\]
 Allowing an apparatus $\A$ to monitor the basis $\BS$, will bring
the $\S\F$ state to Eq. \eqref{eq:rhoSFmu}, where the correlations
will now, by virtue of the conservation law, be given by 
\[
\Hol=\Holp-\QD.
\]
 We see that under monitoring by an apparatus that sees a different
basis, the mutual information is reduced by the discord $\QD$. Thus,
discord with respect to a basis measures the sensitivity of correlations
between the system and fragment.

Despite the reduction of the mutual information between $\S$ and
$\F$ due to the apparatus, the information about the pointer states
that $\S$ deposited in the environment $\E$ is still present in
the fragment $\F$ of that environment. Further, if the apparatus
monitors the pointer state, or close to it, it does not reduce the
mutual information (which is just the classical information). This
gives a consistent picture of what the discord is measuring: It is
measuring the sensitivity of correlations to allowed ``quantum''
monitoring (i.e., monitoring of superposition states) by other environments.
Zero discord in the pointer basis means that another environment can
monitor it with no effect on the $\S$-$\F$ correlations present.
However, when another environment starts to monitor linear combinations
of pointer states, it will reduced the correlations between $\S$
and $\F$.

Going further, for a basis, one can take the definition of discord,
Eq. \eqref{eq:Discord}, and rewrite it as 
\begin{eqnarray*}
\QD & = & \MI-\JI\\
 & = & \MI-I\left(\sum_{s}p_{s}\proj ss\otimes\CS\right)\\
 & = & \MI-I\left(\sum_{s}\sqrt{\pi_{s}}\rho_{\S\F}\sqrt{\pi_{s}}\right),
\end{eqnarray*}
 showing that it is in fact a change of mutual information before
and after a measurement of $\BS$ has been made. One can imagine generalizing
the definition of discord to arbitrary POVMs starting from this expression.
This is not, however, what we have done. Rather, we have used that
$\JI$ is also $\Hol$. This is a more natural generalization in the
context of this paper for two reasons: It separates the mutual information
into two quantities, one representing quantum information and the
other classical. Also, it does not depend on the post-measurement
state of $\S$, which is not uniquely set by the POVM. However, this
more natural generalization is at the expense of interpreting discord
as the loss of correlations under measurement of the POVM. Instead,
discord will be a lower bound to the amount of correlations lost upon
measurement. This is also fitting, since $\Hol$ is the upper bound
to the classically accessible information.

To prove that discord is the lower bound to information lost under
performing the measurement on $\S$, consider an auxiliary system
$\A$ that performs a rank one POVM. In the first step of the measurement,
$\S$ and $\A$ undergo a unitary interaction, with $\A$ initially
decorrelated. This entails that 
\[
\MI=I\left(\A\S^{\p}:\F\right),
\]
 where the prime indicates that the unitary interaction has occurred.
What is also true is that $\Hol=\chi\left(\Pi_{\A\S}:\F\right),$
where $\Pi_{\A\S}$ is the basis on $\A\S$ that results in the rank
one POVM on $\S$. This is because the Holevo quantity only depends
on the conditional states of $\F$ (and its reduced state). Thus,
we have 
\[
\QD=\mathcal{D}\left(\Pi_{\A\S}:\F\right)
\]
 by the conservation law. Using that 
\[
I\left(\sum_{s}\sqrt{\pi_{\A\S,s}}\rho_{\A\S\F}\sqrt{\pi_{\A\S,s}}\right)\ge I\left(\sum_{s}\sqrt{\pi_{s}}\rho_{\S\F}\sqrt{\pi_{s}}\right)
\]
 by the data processing inequality, we have 
\[
\MI-I\left(\sum_{s}\sqrt{\pi_{s}}\rho_{\S\F}\sqrt{\pi_{s}}\right)\ge\QD.
\]
 The left hand side of the inequality depends on how the POVM is performed,
but the right hand side does not. Thus, the inequality holds regardless
of the particulars of the measurement apparatus. This shows that the
generalization of discord using $\Hol$ is independent of the details
of the measurement, and only depends on the POVM.

Let's consider an example of the interpretation of discord for a two
dimensional system. Let $\hat{s}=0,1$ represent the $\sigma_{z}$
basis and $s=\pm$ the $\sigma_{x}$ basis. Starting with surplus
decoherence, the state in the alternative basis $\ket s$ will be
\[
\rho_{\S\F}=\frac{1}{2}\left(\begin{array}{cc}
p_{0}\rho_{\F\left|0\right.}+p_{1}\rho_{\F\left|1\right.} & p_{0}\rho_{\F\left|0\right.}-p_{1}\rho_{\F\left|1\right.}\\
p_{0}\rho_{\F\left|0\right.}-p_{1}\rho_{\F\left|1\right.} & p_{0}\rho_{\F\left|0\right.}+p_{1}\rho_{\F\left|1\right.}
\end{array}\right).
\]
 This shows that, indeed, the $\sigma_{x}$ basis is not classically
correlated with the state of the fragment -- the states on the diagonal
$\rho_{\F\left|+\right.}$ and $\rho_{\F\left|-\right.}$ are the
same. Further, the correlations between the system and fragment have
now been shuffled into the coherences. This latter fact, of course,
does nothing to the correlations between $\S\F$, they are still there.
However, the shuffling gives the immediate -- but not very satisfying
-- interpretation that the correlation is quantum because it is between
a superposition of $+/-$ states and the environment, rather than
those states individually.

More satisfying is that $ $$\QD$ gives the amount of correlations
that would be lost if the system is monitored by an apparatus $\A$
in the $\BS$ basis. Consider $\BS$ as a basis with states $+/-$
rotated by an angle $\mu$ from the pointer basis, as shown in Fig.
\ref{fig:antisym}. After the interaction with the environment $\E$,
if an apparatus $\A$ measures perfectly the $\BS$ basis, then this
will destroy $\QD$ amount of classical correlations with the fragment
$\F$. After tracing out $\A$, the state of $\S\F$ is 
\[
\rho_{\S\F}=p_{+}\proj ++\otimes\rho_{\F\left|+\right.}+p_{-}\proj --\otimes\rho_{\F\left|-\right.},
\]
 with $\rho_{\F\left|+\right.}=p_{0}\rho_{\F\left|0\right.}\cos^{2}\mu/2+p_{1}\rho_{\F\left|1\right.}\sin^{2}\mu/2$
and $\rho_{\F\left|-\right.}=p_{0}\rho_{\F\left|0\right.}\sin^{2}\mu/2+p_{1}\rho_{\F\left|1\right.}\cos^{2}\mu/2$.
That is, the amount of correlations present between $\S$ and $\F$
is now 
\[
\Hol=\Holp-\QD.
\]
 \textcolor{black}{Note that despite the reduction of correlations,
the information $\S$ deposited into $\F$ is still present. The state
of $\F$ gained entropy equal to $H_{\S}$ and this can still be found
out by observers, in which case they will reveal the prior state of
$\S$.}

\end{document}